# Giant interfacial spin-Hall angle from Rashba–Edelstein effect revealed by the spin-Hall Hanle process

Jing Li,[1*] Andrew H. Comstock,[2*] Aeron McConnell,[2] Dali Sun,[2†] and Xiaoshan Xu[1‡]

[1]Department of Physics and Astronomy and Nebraska Center for Materials and Nanoscience, University of Nebraska, Lincoln, Nebraska 68588, USA

[2]Department of Physics and Organic and Carbon Electronics Lab (ORaCEL), North Carolina State University, Raleigh, North Carolina 27695, USA

**Abstract**:

The Rashba–Edelstein effect (REE), which generates interfacial spin polarization and subsequent spin current, is a compelling spin-charge conversion mechanism for spintronic applications, since it's not limited by the elemental spin-orbit couplings. In this work, we demonstrate REE at Pt/ferroelectric interfaces using the recently elucidated spin-Hall Hanle effects (SHHE), in which a Larmor precession of spin polarization in a diffusion process from the interface manifest as magnetoresistance and Hall effect. We show that REE leads to a three-fold enhancement of the effective spin Hall angle in ferroelectric interface Pt/h-LuFeO$_3$ compared to that of Pt/Al$_2$O$_3$, although the difference in the spin relaxation time is negligible. Modeling using SHHEs involving REE as an additional source of interfacial polarization suggests that REE can lead to an interfacial spin Hall angle ($\approx 0.3$) that is one order of magnitude larger than the bulk value of Pt. Our results demonstrate that a ferroelectric interface can produce large spin-charge conversion and that SHHEs are a sensitive tool for characterizing interfacial spin transport properties.

*These authors contributed equally to this work.
†dsun4@ncsu.edu, ‡xiaoshan.xu@unl.edu

# Introduction

Charge-spin current conversion is a fundamental process in spintronics [1]. In general, a charge (spin) current may generate a transverse spin (charge) current via the (inverse) spin Hall effect due to spin-orbit coupling, with a conversion efficiency up to a few percent [2]. With broken inversion symmetry, e.g., at an interface, the gradient of potential energy couples to the spin and generates a preference of spin polarization transverse to the electron momentum [3,4] and subsequent spin current [Fig. 1(a)], corresponding to the Rashba–Edelstein effect (REE) [5,6]. The charge-spin conversion efficiency induced by REE is not limited by the elemental spin-orbit coupling, which is appealing for spintronic applications. In this regard, it is hypothesized that a metal/ferroelectric interface may have greatly enhanced spin-charge conversion efficiency via REE due to the large interfacial potential gradient with the advantage of non-volatile electric-field control [7–10]. On the other hand, determination of REE at metal/ferroelectric interfaces and their spin transport properties, especially at room temperature, has been challenging because of the lack of direct measurements of the spin polarization [11–14].

Here we reveal the presence of REE in Pt/ferroelectric interfaces by employing the spin Hall Hanle effects (SHHEs) technique [15]. SHHE, including Hanle magnetoresistance [16,17] (Hanle MR) and Hanle Hall effect [15], have proven successful in quantifying spin-transport parameters, i.e., spin-Hall angle $\theta_{SH}$, spin diffusion length $\lambda_s$, and spin relaxation time $\tau_s$ in Pt films deposited on non-magnetic insulators which is free from magnetic proximity effects. As illustrated in Fig. 1(b), SHHEs describe the generation of spin current and accumulation of spin polarization at the Pt/insulator interface via the spin Hall effect (SHE [2,18]), precession of spin polarization in the reflected spin current (Hanle effect [19]), and manifestation of spin precession in MR and Hall effects generated by the inverse spin Hall effect (ISHE [20]). Extra interfacial spin accumulation induced by the presence of REE is expected to modify the Hanle MR and Hanle Hall effect, by which REE can be quantitatively studied. We show that the effective spin Hall angle i.e., spin-charge conversion efficiency, $\theta_{SH} \equiv J_c/J_s$, where $J_c$ ($J_s$) is the charge (spin) current density, at the Pt/ferroelectric interface is largely increased compared to that in Pt/Al$_2$O$_3$, while the spin relaxation time remain unchanged. Modelling of the Hanle MR and Hanle Hall effect reveals that REE at the Pt/ferroelectric interface leads to a rescaling of the bulk spin Hall angle. The effective spin-Hall angle of Pt/h-LuFeO$_3$ interface is determined to be one order of magnitude larger than that of bulk Pt.

**Experimental**

We focus on one prototypical ferroelectric family, i.e., hexagonal ferrites (h-$R$FeO$_3$, $R$ = Y, Sc, Ho-Lu) [21,22]. As illustrated in Fig. 1(c), the crystal structure of h-$R$FeO$_3$ consists of layers of FeO$_5$ trigonal bipyramids separated by layers of $R$ ions. With non-centrosymmetric $P6_3cm$ structure, the improper ferroelectricity of h-$R$FeO$_3$ are induced by the K$_3$ structural distortion below ~ 1000 K [22]. Antiferromagnetic orders occur below $T_N$~ 150 K, with Fe spins lying mostly in the basal plane. [23] The K$_3$ distortion can be viewed as the collective tilt of the FeO$_5$ bipyramid and buckling of the $R$ layer, which leads to a moderate spontaneous polarization (≈ 10 μC/cm$^2$) along the $c$ axis [24,25]. Due to the improper nature of ferroelectricity, the spontaneous polarization of h-$R$FeO$_3$ of various $R$ and even that of h-$R$MnO$_3$ (isomorphic to h-$R$FeO$_3$) are about 10 μC/cm$^2$, [24,25] which is the desired material family for introducing the interfacial polarization effect.

Pt(111)/h-LuFeO$_3$(001), Pt(111)/h-YbFeO$_3$(001), and Pt(111)/LuMnO$_3$(001) heterostructures were epitaxially grown on yttrium stabilized zirconia (YSZ) (111) substrates using pulsed laser deposition, with an yttrium aluminum garnet (YAG) laser (266 nm wavelength, 70 mJ pulse energy, 3 Hz repetition rate). [See Supplementary materials S1] The thickness of h-LuFeO$_3$, h-YbFeO$_3$, and LuMnO$_3$ layers are about 15 nm which were deposited in 20 mTorr O$_2$ with 650 °C substrate temperature [23,26], followed by the deposition of the Pt layer of various thickness in $10^{-7}$ Torr vacuum at room temperature without breaking the vacuum. The Pt layer was patterned into Hall bar by photolithography and ion milling [15]. The thicknesses of thin films were measured using x-ray reflectivity. Longitudinal ($\rho_L$) and transverse ($\rho_T$) resistivity was measured using the Hall bar configuration in magnetic field applied along different directions at room temperature [15]. Field dependence of $\rho_L$ and $\rho_T$ was symmetrized and antisymmetrized, respectively, to minimize the possible artifacts due to sample misalignment with the field.

## Results and Discussion

Figure 2 shows the obtained MR (i.e., field dependence of $\rho_L$) and the Hall effect (i.e., field dependence of $\rho_T$) in the Pt (5.4 nm)/h-LuFeO$_3$ and the control sample, Pt (5.2 nm)/Al$_2$O$_3$ [15]. The MRs in both samples show a symmetry curve that is consistent with reported SHHE in $\rho_L$, i.e., MR($B_z$) and MR($B_x$) are almost identical and much larger than MR($B_y$), similar to that of spin Hall magnetoresistance [27,28] but without the presence of magnetic materials. As illustrated in Fig. 1(b), SHHEs include contributions from SHE, Hanle effect, and ISHE: a charge current flowing along the *x* direction produces a spin current along the *z* direction toward the Pt/insulator interface with spin polarization along the y direction via SHE; the precession of spin polarization of the reflected spin current in a magnetic field during the spin diffusion reduces the *y* component (Hanle effect) and generates the *x* or *z* component of the spin polarization; the spin precession shows up in both MR and Hall effect since the reflected spin current generates charge current via the ISHE.

Hence, MR($B_x$) and MR($B_z$) are expected to be larger than MR($B_y$) since $B_y$ does not lead to the spin precession. Remarkably, we found that the magnitude of both MR and Hall effect are much larger in Pt(5.4 nm)/h-LuFeO$_3$ than that in Pt(5.2 nm)/Al$_2$O$_3$, implying that the spin Hall angle $\theta_{SH}$ is substantially larger in Pt/h-LuFeO$_3$.

To determine $\theta_{SH}$ in Pt/h-LuFeO$_3$, the spin diffusion $\lambda_s$ length is needed.[15] Essentially, SHHEs vanish in both thin-film ($d/\lambda_s \to 0$) and thick-film limits ($d/\lambda_s \to \infty$). Therefore, $\lambda_s$ can be extracted from the thickness dependence of SHHE signal. By applying $B_z$, SHHEs are described using the following equation [15]

$$\frac{\Delta \rho_{SHHE}}{\rho_{L0}} = \theta_{SH}^2 \frac{\tanh(d/2\lambda_s)}{d/2\lambda_s}\left[1 - \frac{1}{\kappa}\frac{\tanh\left(\frac{\kappa d}{2\lambda_s}\right)}{\tanh\left(\frac{d}{2\lambda_s}\right)}\right], \tag{1}$$

where $\rho_{L0}$ is zero-field longitudinal resistivity, $d$ is the film thickness, $\kappa = (1 - i\Omega\tau_s)^{1/2}$ is a complex quantity with $i = \sqrt{-1}$, $\Omega = g\mu_B B_z/\hbar$ is the Larmor frequency with $g$ the gyromagnetic factor, $\mu_B$ the Bohr magneton, and $\hbar$ the reduced Planck constant. The real and imaginary parts of $\Delta\rho_{SHHE}$ represent the longitudinal $\Delta\rho_{L,SHHE}$ and the transverse $\rho_{T,SHHE}$, respectively. It is shown that at low field regime, $\rho_{T,SHHE}$ would be linear to $\Omega$ (or $B_z$) [15]. By fitting the slope of $\rho_T(B_z)$ at low field regime [Fig. 2(c)], we found the spin diffusion length as $\lambda_s = 0.9 \pm 0.2$ nm for Pt/h-LuFeO$_3$.

Once $\lambda_s$ is determined, the spin Hall angle $\theta_{SH}$ and spin relaxation time $\tau_s$ can be extracted subsequently from the MR $\Delta\rho_L(B_z)$ and Hall effect $\rho_T(B_z)$ simultaneously. As shown in Fig. 3(a) and (b), the measured MR and Hall effect can be well fitted using Eq. (1). It is noteworthy that for each sample, the fit for both MR and Hall effect uses the same set of parameters $\theta_{SH}$ and $\tau_s$. The extracted $\theta_{SH}$ and $\tau_s$ are displayed in Fig. 4. The obtained $\theta_{SH}$ of Pt grown on h-LuFeO$_3$ is roughly 3 times larger than that of Pt grown on Al$_2$O$_3$ [Fig. 4(a)], whereas $\tau_s$ in both two systems are quite similar [Fig. 4(b)].

To further validate that the enhancement of θ$_{SH}$ is induced by the ferroelectric interface of Pt/h-LuFeO$_3$, we conducted similar SHHE measurements in Pt/h-YbFeO$_3$ and Pt/LuMnO$_3$ at different Pt thicknesses. LuMnO$_3$ and h-YbFeO$_3$ are both isomorphic to h-LuFeO$_3$, with the same crystal structure and lattice distortion that result in the improper ferroelectricity. Although the rare earth-sites and the transition metal sites are substituted by different elements in h-YbFeO$_3$ and LuMnO$_3$ respectively, the spontaneous polarizations in all three materials are roughly the same.

The MR Δρ$_L$($B_z$) and Hall effect ρ$_T$($B_z$) of Pt/h-YbFeO$_3$ and Pt/LuMnO$_3$ are shown in Fig. 3(c) and (d). All the curves can be well fitted using Eq. (1); the fitting parameters are displayed in Fig. 4. Obtained θ$_{SH}$ and τ$_s$ of Pt/h-YbFeO$_3$ and Pt/LuMnO$_3$ are consistent with that of Pt/h-LuFeO$_3$. We found the averaged value of θ$_{SH}$ = 0.06 ± 0.01 and τ$_s$ = 1.8 ± 0.4 ps in Pt/h-$R$FeO$_3$ as displayed in Table I. There is a three times enhancement of θ$_{SH}$ in Pt/h-$R$FeO$_3$ system compared with that in Pt/Al$_2$O$_3$ with negligible difference in τ$_s$. Considering the similarity in the spontaneous polarization but large differences in composition at the three interfaces, the enhanced θ$_{SH}$ is attributed to the presence of the ferroelectric interface.

**Model**

Below we provide a model in which REE at the Pt/ferroelectric interface can lead to the enhancement of θ$_{SH}$. At the Pt/ferroelectric interface, the large electric potential difference and spin-orbit coupling adds a Rashba term in the Hamiltonian of the itinerant electrons, i.e., $\widehat{H}_R \propto \vec{\sigma} \cdot (\vec{p} \times \hat{z})$, where $\vec{\sigma}$, $\vec{p}$, and $\hat{z}$ are the Pauli matrices representing the electronic spin, momentum, and surface normal unit vector [5]. For a charge current with an average momentum $\langle \vec{p} \rangle$ along the $x$ direction, $\langle \vec{p} \rangle \times \hat{z}$ acts like an effective magnetic field along the $y$ direction [5]. This mechanism (or REE) is an extra source of spin accumulation along the $y$ axis at the interface [Fig. 1(a)] which can contribute to the spin current along the $z$ direction via subsequent diffusion [12–14].

To model the extra spin accumulation via REE, we use the following equations that govern the spin transport at the Pt/ferroelectric interface [29,30]:

$$q_i = -\mu E_i + \theta_{SH}\epsilon_{ijk}q_{jk} \quad (2)$$

$$q_{ij} = -\mu E_i P_j - D\frac{\partial P_j}{\partial x_i} - \theta_{SH}\epsilon_{ijk}q_k \quad (3)$$

$$\epsilon_{ijk}P_j\Omega_k + \frac{\partial q_{ki}}{\partial x_k} + \frac{P_i}{\tau_s} - \frac{\beta}{L_R}(\vec{q}\times\hat{z})_i e^{-\frac{2z+d}{2L_R}} = 0 \quad (4)$$

where $q_i$ and $q_{ij}$ ($i, j = x, y, z$) are the components of the charge and spin current density respectively in (unified) units of particle per unit time per unit area [29], $x_i$, $\Omega_i$, $P_i$ and $E_i$ are the $i$th component of spatial coordinate, Larmor precession $\vec{\Omega} = g\mu_B\vec{B}/\hbar$, spin polarization density, and electric field respectively, $d$, $\mu$, $D$ and $\epsilon_{ijk}$ are film thickness, mobility times carrier density, diffusion coefficient, and antisymmetric matrix respectively.

Eqs. (2) and (3) are essentially the "Ohm's law" for charge and spin currents, including the typical drift and diffusion processes, as well as SHE and ISHE [29]. Eq. (4) ensures the continuity of spin polarization. The term $\frac{\beta}{L_R}(\vec{q}\times\hat{z})_i e^{-\frac{2z+d}{2L_R}}$ is introduced to describe the source of spin polarization at the interface ($z = d/2$) due to REE, where $\beta$ is a dimensionless coefficient. $L_R$ is the nominal thickness of the interfacial layer which is presumably much smaller than the film thickness $d$ and spin diffusion length $\lambda_s$ for Pt.

Solving Eqs. (2-4) [see Supplementary Material S2] with the boundary conditions $\frac{\partial P_i}{\partial x} = \frac{\partial P_i}{\partial z} = 0$ and $q_{zi}\left(z = -\frac{d}{2}, \frac{d}{2}\right) = 0$, we found

$$\frac{\rho_{SHHE}}{\rho_0} = \theta_{SH}^2\left(1 + \frac{1}{2}\frac{\beta}{\theta_{SH}}\right)\frac{\tanh\left(\frac{d}{2\lambda_s}\right)}{\frac{d}{2\lambda_s}}\left[1 - \frac{1}{\kappa}\frac{\tanh\left(\frac{\kappa d}{2\lambda_s}\right)}{\tanh\left(\frac{d}{2\lambda_s}\right)}\right]. \quad \text{Eq. (5)}$$

Eq. (5) differs from Eq. (1) only by a rescaling of $\theta_{SH}^2 \rightarrow \theta_{SH}^2 \left(1 + \frac{1}{2}\frac{\beta}{\theta_{SH}}\right)$. In other words, the interfacial Rashba effect can manifest as an enhancement of effective bulk spin Hall angle $\theta_{SH}$. Notice that integration of $\frac{\beta}{L_R}(\vec{q} \times \hat{z})_i e^{-\frac{2z+d}{2L_R}}$ over the entire film thickness ($z = 0 - d$) leads to a spin current $\beta q$, suggesting that $\beta$ is the interfacial spin Hall angle originating from REE. Using $\theta_{SH}$ = 0.022±0.006 measured from Pt/Al$_2$O$_3$ [15] and $\theta_{SH}\sqrt{1 + \frac{1}{2}\frac{\beta}{\theta_{SH}}}$ = 0.06 ± 0.01 measured in Pt deposited on hexagonal ferrites and manganites (see Table I), we find $\beta$ = 0.3 ± 0.1, which is about an order of magnitude larger than the bulk spin Hall angle $\theta_{SH}$ in Pt.

## Conclusion

SHHEs are utilized to study the spin accumulation at the Pt/ferroelectric interface due to the Rashba–Edelstein effect. A three-fold enhancement of effective spin Hall angle was observed compared to that in Pt/Al$_2$O$_3$ whereas the spin relaxation times remain unchanged, corroborated by the modeling using the spin Hall Hanle process involving the Rashba–Edelstein effect as additional source of interfacial spin accumulation. The effective interfacial spin Hall angle was extracted as one order of magnitude larger than the bulk value. These results suggest that ferroelectric interfaces are promising for efficient spin-charge interconversion for future spintronic applications.

## Acknowledgements

This research was primarily supported by the U.S. Department of Energy (DOE), Office of Science, Basic Energy Sciences (BES), under Award No. DE-SC0019173. The work at NC State was supported by the U.S. DOE, Office of Science, BES, under Award No. DE-SC0020992. The research was performed in part in the Nebraska Nanoscale Facility: National Nanotechnology Coordinated Infrastructure and the Nebraska Center for Materials and Nanoscience (and/or NERCF), which are supported by the National Science Foundation under Award No. ECCS: 1542182, and the Nebraska Research Initiative.

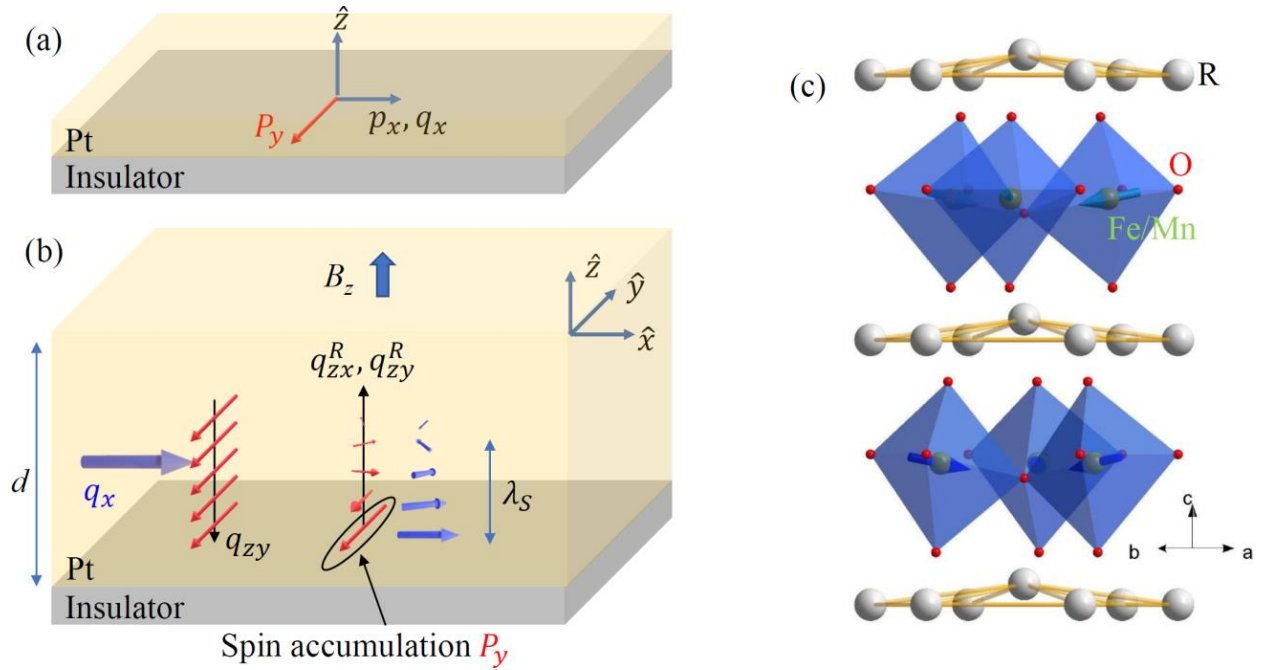

**Figure 1.** (a) Schematic of the Rashba–Edelstein effect (REE) in which a charge current at the interface leads to a transverse spin polarization. (b) Schematic of the spin Hall Hanle effects (SHHEs) in a Pt/insulator heterostructure. Applied magnetic field causes precession of spin polarization during the diffusion from the interface and manifests in the magnetoresistance and the Hall effect. (c) A unit cell of ferroelectric hexagonal ferrites (manganites) featuring $FeO_5$ ($MnO_5$) layers separated by the rare earth layers. Arrows through Fe atoms represent spin orientations.

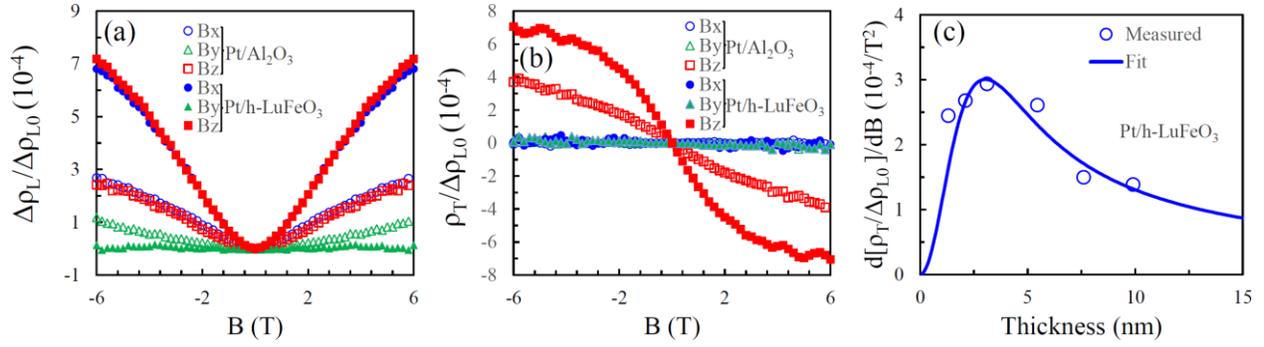

**Figure 2.** Field dependence of the normalized longitudinal resistivity (magnetoresistance) (a) and transverse resistivity (Hall effect) of a Pt (5.4 nm)/h-LuFeO$_3$ and a Pt (5.2 nm)/Al$_2$O$_3$ samples. (c) Thickness dependence of the low-field slope of the Hall effect for Pt/h-LuFeO$_3$. The fitting reveals a spin diffusion length $\lambda_s = 0.9 \pm 0.2$ nm.

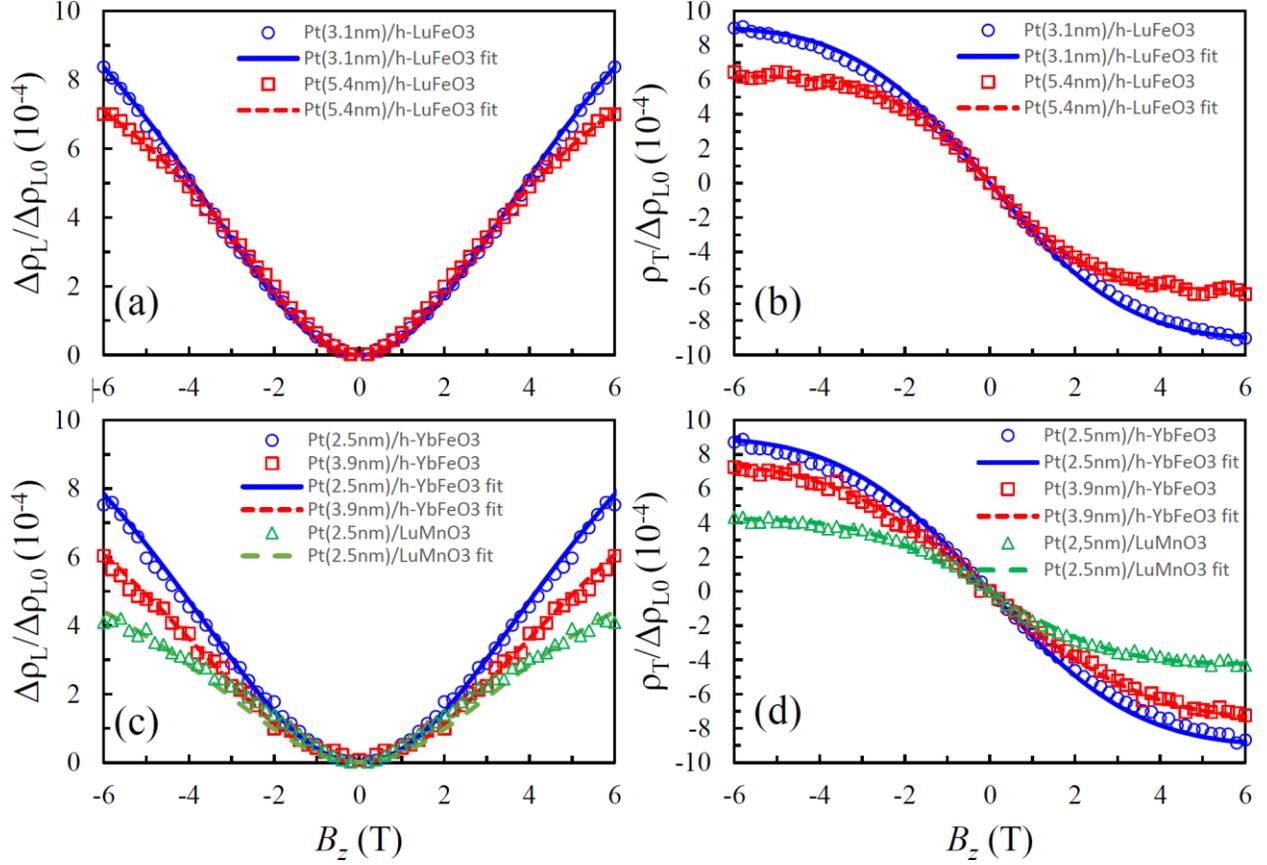

**Figure 3**. (a) and (c) are the measured MR (symbol) as the difference between the magnetoresistance in $B_z$ and that in $B_y$. (b) and (d) are the measured Hall effect (symbol). The lines in are fits of the data using the Hanle MR and the Hanle Hall effect according to Eq. (1). For each film, the fit for both MR and Hall effect uses the same set of parameters.

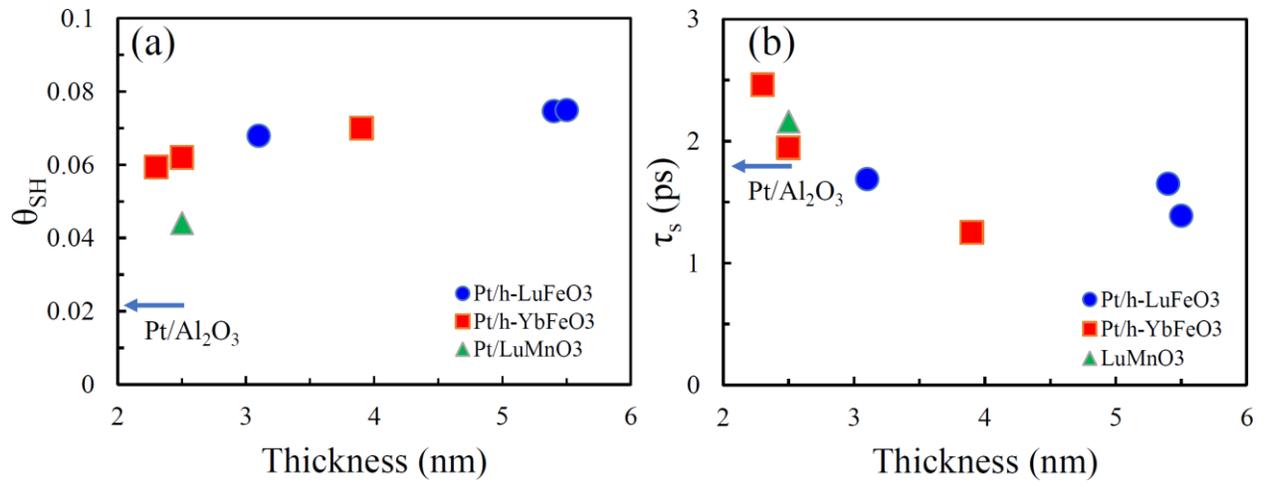

**Figure 4**. Spin Hall angle $\theta_{SH}$ (a) and spin relaxation time $\tau_s$ (b) extracted from the fitting in Fig. (3).

Table I. Comparison of spin transport parameters of Pt/$Al_2O_3$ and that of Pt/hexagonal ferrites and Pt/manganite combined.

|  | Pt/$Al_2O_3$ | Pt/h-$LuFeO_3$ Pt/h-$YbFeO_3$ Pt/$LuMnO_3$ |
|---|---|---|
| Spin-Hall angle | $0.022 \pm 0.006$ | $0.06 \pm 0.01$ |
| Spin diffusion length (nm) | $1.63 \pm 0.26$ nm | $0.9 \pm 0.2$ |
| Spin relaxation time (ps) | $1.8 \pm 0.9$ | $1.8 \pm 0.4$ |

Supplementary Materials:

Giant interfacial spin-Hall angle from Rashba–Edelstein effect revealed by the spin-Hall Hanle process

Jing Li,[1] Andrew H. Comstock,[2] Dali Sun,[2] and Xiaoshan Xu[1]

[1]Department of Physics and Astronomy and Nebraska Center for Materials and Nanoscience, University of Nebraska, Lincoln, Nebraska 68588, USA

[2]Department of Physics and Organic and Carbon Electronics Lab (ORaCEL), North Carolina State University, Raleigh, North Carolina 27695, USA

## S1. X-ray diffraction

X-ray diffraction (XRD) and x-ray rocking curve of Pt films on h-LuFeO$_3$ were measured using Rigaku SmartLab Diffractometer with Cu K$\alpha$ radiation (wavelength 1.54 Angstrom). As shown in Fig. S1(a), the h-LuFeO$_3$ (002) peak in this specular diffraction indicates epitaxial growth with *c* axis along the out-of-plane direction. The narrow rocking curve suggests high crystallinity of the h-LuFeO$_3$ film.

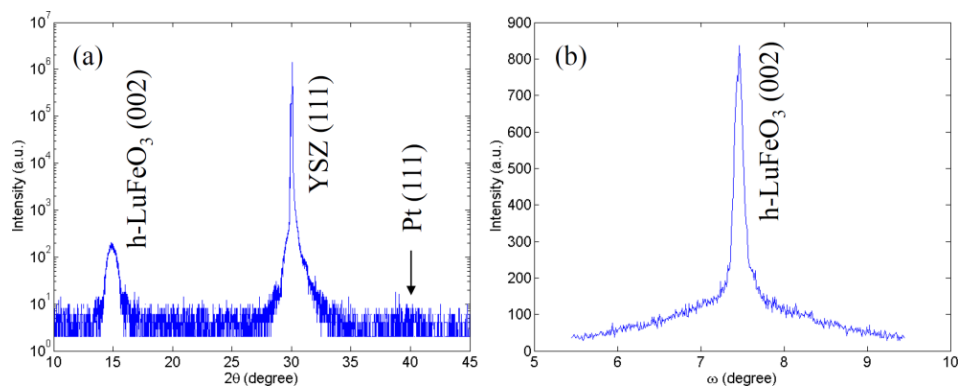

**Fig. S1** X-ray diffraction pattern (a) and rocking curve (b) of a Pt (3.1 nm)/h-LuFeO$_3$ (15 nm)/YSZ sample.

## S2. Derivation of Spin-Hall Hanle effects with contribution from the Rashba–Edelstein effect

2.1 Equation of motion

$$q_i = -\mu E_i + \gamma \epsilon_{ijk} q_{jk}$$

$$q_{ij} = -\mu E_i P_j - D \frac{\partial P_j}{\partial x_i} - \gamma \epsilon_{ijk} q_k$$

$$\epsilon_{ijk} P_j \Omega_k + \frac{\partial q_{ki}}{\partial x_k} + \frac{P_i}{\tau_s} - \frac{\beta}{L_R}(\vec{q} \times \hat{z})_i e^{-\frac{2z+d}{2L_R}} = 0.$$

$q_i$ and $q_{ij}$ ($i,j = x, y, z$) are charge and spin current unified in particle current density. The "normal" charge and spin current is given by $eq_i$ and $\hbar q_{ij}$ respectively. $\gamma$ and $\tau_s$ are the spin Hall angle and spin relaxation time. $x_i$, $\Omega_i$, $P_i$ and $E_i$ are the $i$th component of spatial coordinate, Larmor precession $\vec{\Omega} = g\mu_B \vec{B}/\hbar$, spin polarization density, and electric field respectively, $d$, $\mu$, $D$ and $\epsilon_{ijk}$ are film thickness, mobility times carrier density, diffusion coefficient, and antisymmetric matrix respectively.

We basically add a "source" of the spin polarization $\frac{\beta}{L_R}(\vec{q} \times \hat{z})_i e^{-\frac{2z+d}{2L_R}}$ at the interface $z = -d/2$ with thickness $L_R$. Since $\int_{-\frac{d}{2}}^{\frac{d}{2}} \frac{1}{L_R} e^{-\frac{2z+d}{2L_R}} dz = e^{-\frac{d}{2L_R}} \int_{-\frac{d}{2L_R}}^{\frac{d}{2L_R}} e^{-\eta} d\eta = e^{-\frac{d}{2L_R}}\left(e^{\frac{d}{2L_R}} - e^{-\frac{d}{2L_R}}\right) = 1 - e^{-\frac{d}{L_R}} \approx 1$ when $d \gg L_R$, the dimensionless parameter $\beta$ represents the strength of the Rashba-Edelstein effect. Given $\vec{q}$ is only along the $x$ direction, this source will only contribute to $P_y$.

2.2 Solution

Considering the sample geometry and follow the steps:

Step (1) charge current

$$q_x \approx -\mu E_x = q_{x0}, q_y \approx 0, q_z \approx 0$$

Step (2) spin current

$$q_{xx} = -\mu E_x P_x, q_{xy} = -\mu E_x P_y, q_{xz} = -\mu E_x P_z$$

$$q_{yx} = 0, q_{yy} = 0, q_{yz} = 0$$

$$q_{zx} = -D\frac{\partial P_x}{\partial z}, q_{zy} = -D\frac{\partial P_y}{\partial z} + \gamma q_x, q_{zz} = -D\frac{\partial P_z}{\partial z}$$

Step (3) continuity equation

$$P_y \Omega_z + \frac{\partial q_{zx}}{\partial z} + \frac{P_x}{\tau_s} = 0$$

$$-P_x \Omega_z + \frac{\partial q_{zy}}{\partial z} + \frac{P_y}{\tau_s} + \frac{\beta}{L_R} q_x e^{-\frac{2z+d}{2L_R}} = 0$$

$$\frac{\partial q_{zz}}{\partial z} + \frac{P_z}{\tau_s} = 0$$

Plug in the expression for the spin currents, one has

$$P_y \Omega_z - D\frac{\partial^2 P_x}{\partial z^2} + \frac{P_x}{\tau_s} = 0$$

$$-P_x \Omega_z - D\frac{\partial^2 P_y}{\partial z^2} + \frac{P_y}{\tau_s} + \frac{\beta}{L_R} q_x e^{-\frac{2z+d}{2L_R}} = 0$$

$$-D\frac{\partial^2 P_z}{\partial z^2} + \frac{P_z}{\tau_s} = 0$$

2.2.1 $P_z$

This can be solved directly as

$$P_z = A e^{\frac{z}{\lambda_s}} + B e^{-\frac{z}{\lambda_s}},$$

where $\lambda_s = \sqrt{D\tau_s}$ is the spin diffusion length.

The corresponding spin current is

$$q_{zz} = -D\frac{\partial P_z}{\partial z} = -D\left(-\frac{A}{\lambda_s} e^{\frac{z}{\lambda_s}} + \frac{B}{\lambda_s} e^{-\frac{z}{\lambda_s}}\right)$$

The boundary condition is $q_{zz}\left(\frac{d}{2}\right) = q_{zz}\left(-\frac{d}{2}\right) = 0$, which means

$$-D\left(-\frac{A}{\lambda_s} e^{\frac{d}{2\lambda_s}} + \frac{B}{\lambda_s} e^{-\frac{d}{2\lambda_s}}\right) = 0 \rightarrow A = B e^{-\frac{d}{\lambda_s}}$$

$$-D\left(-\frac{A}{\lambda_s} e^{-\frac{d}{2\lambda_s}} + \frac{B}{\lambda_s} e^{\frac{d}{2\lambda_s}}\right) = 0 \rightarrow A = B e^{\frac{d}{\lambda_s}}$$

This means $A = B = 0$.

2.2.2 $P_y$

*2.2.2.1 Spin polarization*
For the x-y directions, one has

$$P_y \Omega_z - D\frac{\partial^2 P_x}{\partial z^2} + \frac{P_x}{\tau_s} = 0$$

$$-P_x \Omega_z - D\frac{\partial^2 P_y}{\partial z^2} + \frac{P_y}{\tau_s} + \frac{\beta}{L_R} q_{x0} e^{-\frac{2z+d}{2L_R}} = 0.$$

We solve from the 2$^{nd}$ equation above,

$$P_x = \frac{1}{\Omega_z}\left(-D\frac{\partial^2 P_y}{\partial z^2} + \frac{P_y}{\tau_s} + \frac{\beta}{L_R}q_{x0}e^{-\frac{2z+d}{2L_R}}\right)$$

Plug into the 1st equation above, we get

$$\frac{\partial^4 P_y}{\partial z^4} - 2\frac{1}{D\tau_s}\frac{\partial^2 P_y}{\partial z^2} + \frac{1}{D^2\tau_s^2}(1+\Omega_z^2\tau_s^2)P_y = -\left(\frac{1}{\tau_s D} - \frac{1}{L_R^2}\right)\frac{\beta}{DL_R}q_{x0}e^{-\frac{2z+d}{2L_R}}.$$

With the boundary conditions $q_{zy}\left(\frac{d}{2}\right) = q_{zy}\left(-\frac{d}{2}\right) = 0$, we find

$$P_y = \gamma\frac{\lambda_+}{\lambda_s}P_0\frac{e^{\frac{z}{\Lambda}} - e^{-\frac{z}{\Lambda}}}{e^{\frac{d}{2\Lambda}} + e^{-\frac{d}{2\Lambda}}} - \beta\left(1-e^{-\frac{d}{L_R}}\right)\frac{\Lambda}{\lambda_s}P_0\frac{e^{\frac{2z-d}{2\Lambda}} + e^{-\frac{2z-d}{2\Lambda}}}{e^{\frac{d}{\Lambda}} - e^{-\frac{d}{\Lambda}}} + \frac{\beta\lambda_+}{\lambda_s}P_0 e^{-\frac{d}{L_R}}\frac{e^{\frac{z}{\Lambda}} - e^{-\frac{z}{\Lambda}}}{e^{\frac{d}{2\Lambda}} + e^{-\frac{d}{2\Lambda}}}$$
$$+ \beta\frac{L_R}{\lambda_s}P_0 e^{-\frac{2z+d}{2L_R}}$$

where $P_0 \equiv \frac{\lambda_s}{D}q_{x0}$, $\tau_s D = \lambda_s^2 \gg L_R^2$, $\Lambda \equiv \frac{\lambda_s}{\sqrt{1-j\Omega\tau_s}} = \frac{\lambda_s}{k}$, $k = \sqrt{1-j\Omega\tau_s}$.

*2.2.2.2 Spin current*
The corresponding spin current is

$$q_{zy} = -D\frac{\partial P_y}{\partial z} + \gamma q_{x0}$$

$$= \gamma q_{x0}\left[1 - \frac{e^{\frac{z}{\Lambda}} + e^{-\frac{z}{\Lambda}}}{e^{\frac{d}{2\Lambda}} + e^{-\frac{d}{2\Lambda}}} + \frac{\beta}{\gamma}\left(1-e^{-\frac{d}{L_R}}\right)\frac{e^{\frac{2z-d}{2\Lambda}} - e^{-\frac{2z-d}{2\Lambda}}}{e^{\frac{d}{\Lambda}} - e^{-\frac{d}{\Lambda}}} - \frac{\beta}{\gamma}e^{-\frac{d}{L_R}}\frac{e^{\frac{z}{\Lambda}} + e^{-\frac{z}{\Lambda}}}{e^{\frac{d}{2\Lambda}} + e^{-\frac{d}{2\Lambda}}} + \frac{\beta}{\gamma}e^{-\frac{2z+d}{2L_R}}\right]$$

where $P_0 \equiv \frac{\lambda_s}{D}q_{x0}$ is used.

*2.2.2.3 Charge current from ISHE and MR*
The corresponding charge current is

$$q'_x(B,z) = -\gamma q_{zy}$$

$$= -\gamma^2 q_{x0}\left[1 - \frac{e^{\frac{z}{\Lambda}} + e^{-\frac{z}{\Lambda}}}{e^{\frac{d}{2\Lambda}} + e^{-\frac{d}{2\Lambda}}} + \frac{\beta}{\gamma}\left(1-e^{-\frac{d}{L_R}}\right)\frac{e^{\frac{2z-d}{2\Lambda}} - e^{-\frac{2z-d}{2\Lambda}}}{e^{\frac{d}{\Lambda}} - e^{-\frac{d}{\Lambda}}} - \frac{\beta}{\gamma}e^{-\frac{d}{L_R}}\frac{e^{\frac{z}{\Lambda}} + e^{-\frac{z}{\Lambda}}}{e^{\frac{d}{2\Lambda}} + e^{-\frac{d}{2\Lambda}}} + \frac{\beta}{\gamma}e^{-\frac{2z+d}{2L_R}}\right]$$

Integrate over z = (-d/2, d/2),

$$q'_x(B) = -\gamma^2 q_{x0}\left[1 - \left(1 + \frac{1}{2}\frac{\beta}{\gamma} - \frac{1}{2}\frac{\beta}{\gamma}e^{-\frac{d}{L_R}}\right)\frac{2\lambda_+}{d}\frac{e^{\frac{d}{2\Lambda}} - e^{-\frac{d}{2\Lambda}}}{e^{\frac{d}{2\Lambda}} + e^{-\frac{d}{2\Lambda}}} + \frac{\beta}{\gamma}\frac{L_R}{d}\left(1 - e^{-\frac{d}{L_R}}\right)\right]$$

When the film is not too thin, i.e., $d \gg L_R$, $\frac{L_R}{d} \to 0$, $e^{-\frac{d}{L_R}} \to 0$, one can ignore the last term. This is simplified as

$$q'_x(B) \approx \gamma^2 q_{x0} \left[1 - \left(1 + \frac{1}{2}\frac{\beta}{\gamma}\right) \frac{2\Lambda}{d} \frac{e^{\frac{d}{2\Lambda}} - e^{-\frac{d}{2\Lambda}}}{e^{\frac{d}{2\Lambda}} + e^{-\frac{d}{2\Lambda}}}\right].$$

We can calculate MR as

$$\text{MR} = -\frac{q'_x(B) - q'_x(B=0)}{q_x} = \gamma^2 \left(1 + \frac{1}{2}\frac{\beta}{\gamma}\right) \left(\frac{2\lambda_s}{d} \frac{e^{\frac{d}{2\lambda_s}} - e^{-\frac{d}{2\lambda_s}}}{e^{\frac{d}{2\lambda_s}} + e^{-\frac{d}{2\lambda_s}}} - \frac{2\lambda_+}{d} \frac{e^{\frac{d}{2\Lambda}} - e^{-\frac{d}{2\Lambda}}}{e^{\frac{d}{2\Lambda}} + e^{-\frac{d}{2\Lambda}}}\right)$$

$$= \gamma^2 \left(1 + \frac{1}{2}\frac{\beta}{\gamma}\right) \frac{\tanh\left(\frac{d}{2\lambda_s}\right)}{\frac{d}{2\lambda_s}} \left[1 - \frac{\Lambda}{\lambda_s} \tanh\left(\frac{d}{2\Lambda}\right)\right]$$

$$= \gamma^2 \left(1 + \frac{1}{2}\frac{\beta}{\gamma}\right) \frac{\tanh\left(\frac{d}{2\lambda_s}\right)}{\frac{d}{2\lambda_s}} \left[1 - \frac{1}{\kappa} \frac{\tanh\left(\frac{\kappa d}{2\lambda_s}\right)}{\tanh\left(\frac{d}{2\lambda_s}\right)}\right]$$

### 1.2.3 $P_x$
*2.2.3.1 Spin polarization*

We find $P_x$ using

$$P_x = \frac{1}{\Omega_z}\left(-D\frac{\partial^2 P_y}{\partial z^2} + \frac{P_y}{\tau_s} + \frac{\beta}{L_R} q_{x0} e^{-\frac{2z+d}{2L_R}}\right)$$

From the expression of $P_y$.

We consider $\frac{L_R}{d} \ll 1$ and $\frac{L_R}{\lambda_s} \ll 1$. One has

$$P_x \approx -j\left(\gamma \frac{\Lambda}{\lambda_s} P_0 \frac{e^{\frac{z}{\Lambda}} - e^{-\frac{z}{\Lambda}}}{e^{\frac{d}{2\Lambda}} + e^{-\frac{d}{2\Lambda}}} - \beta \frac{\Lambda}{\lambda_s} P_0 \frac{e^{\frac{z-2d}{2\Lambda}} - e^{-\frac{z-2d}{2\Lambda}}}{e^{\frac{d}{\Lambda}} - e^{-\frac{d}{\Lambda}}}\right).$$

Compared with

$$P_y = \gamma \frac{\Lambda}{\lambda_s} P_0 \frac{e^{\frac{z}{\Lambda}} - e^{-\frac{z}{\Lambda}}}{e^{\frac{d}{2\Lambda}} + e^{-\frac{d}{2\Lambda}}} - \beta \frac{\lambda_+}{\lambda_s} P_0 \frac{e^{\frac{2z-d}{2\Lambda}} + e^{-\frac{2z-d}{2\Lambda}}}{e^{\frac{d}{\Lambda}} - e^{-\frac{d}{\Lambda}}} + \beta \frac{L_R}{\lambda_s} P_0 e^{-\frac{2z+d}{2L_R}}$$

Here $P_x$ is the imaginary part of field-dependent part of $P_y$.

*2.2.3.2 Spin current*

One can tell that the field dependent part of $P_x$ and $P_y$ are imaginary and real parts of the same complex value, which is the same case as that of the no Rashba effects.

In the end, we will get a similar effect like that in SHHE but with a rescaled spin Hall angle

$$\frac{\rho_{SHHE}}{\rho_0} = \gamma^2 \left(1 + \frac{1}{2}\frac{\beta}{\gamma}\right) \frac{\tanh\left(\frac{d}{2\lambda_s}\right)}{\frac{d}{2\lambda_s}} \left[1 - \frac{1}{\kappa}\frac{\tanh\left(\frac{\kappa d}{2\lambda_s}\right)}{\tanh\left(\frac{d}{2\lambda_s}\right)}\right]$$

Essentially, now $\gamma^2$ is replaced with $\gamma^2 \left(1 + \frac{1}{2}\frac{\beta}{\gamma}\right)$.